%% file: dissa.tex
\documentclass[aps,prl,twocolumn,groupaddress,floats,showpacs]{revtex4}
\usepackage{amsfonts}
\usepackage{amssymb}
\usepackage{bm}
\usepackage[final]{graphicx}

\begin{document}
\bibliographystyle{apsrev}

\include{defs}

\title{Spin asymmetries for confined Dirac particles}

\author{Vijay R.\ Pandharipande}
\email[]{vrp@uiuc.edu}
\affiliation{Department of Physics,
University of Illinois at Urbana-Champaign,
Urbana, Illinois 61801}
\author{Mark W.\ Paris}
\email[]{paris@lanl.gov}
\affiliation{Theoretical Division,
Los Alamos National Laboratory,
Los Alamos, NM 87545}
\author{Ingo Sick}
\email[]{Ingo.Sick@unibas.ch}
\affiliation{Departement f\"{u}r Physik und Astronomie,
Universit\"{a}t Basel, Switzerland}

\date{\today}

\begin{abstract}
\medskip
We study the Bjorken $x$ (or equivalently Nachtmann $\xi$) dependence of
the virtual photon spin asymmetry in polarized deep inelastic scattering
of electrons from hadrons. We use an exactly solved relativistic potential model of 
the hadron, treating the constituents as independent massless Dirac particles
confined to an infinitely massive force center. The importance of including 
the $p$--wave components of the Dirac wave function is demonstrated. 
Comparisons are made to the observed data on the proton by taking into
account the observed flavor dependence of the valence quark distribution 
functions.
\end{abstract}
\maketitle

The structure functions of the nucleon as measured by 
deep inelastic scattering (DIS) of leptons from nucleons have, 
over the last 30 years, received much attention. In particular,
the evolution with momentum transfer $Q^2$ has been quantitatively
understood in terms of quantum chromodynamics (QCD).

During the last decade, the interest has been concentrated on the 
{\em spin structure functions}. Much of this interest was due to the 
fact that the integral over the experimental spin structure function 
$g_1 (x)$ yielded values that were much lower than the ones expected 
in the naive quark model 
(for reviews see e.g. \cite{Filippone01,Anselmino95,Hughes99}).
The presence of this ``spin crisis''  has led to many different 
ideas on how to account for  the nucleon spin. The contribution 
of $s\bar{s}$-components, the contribution of orbital angular momenta of 
the quarks and the contribution of gluons have been discussed. 
It also has been pointed out early on \cite{Bjorken66} that the 
non-relativistic quark model overestimates the quark contribution
to the nucleon spin. Relativistic effects lead to a reduction
which today, together  with the gluon contribution via the 
triangle diagram, are believed to provide the main explanation 
for the low integral over the spin structure function $g_1$; 
in the present letter, we concentrate on the former.

Relativistic effects are expected to play an important role as 
the masses of quarks are small compared with their momenta. 
The non-relativistic quark 
models for instance also overestimate the axial vector weak 
coupling constants which experimentally amount to $g_A/g_V = 1.26$ 
rather than $5/3$. Calculations with the MIT bag model \cite{Jaffe90} 
or using light-cone quantization \cite{Brodsky94} have indicated 
that the lower components of the wave function --- present in 
a relativistic description using the Dirac equation --- lead to 
an opposite contribution to the one of the upper components, 
which could generate, in the limit of massless quarks, a reduction 
factor of 0.65 \cite{JaffeManohar90}.

In the present Letter we study a simple model for relativistic 
bound quarks. We want to address in particular the $x$-dependence 
of $g_1(x)$, for which otherwise little  theoretical guidance is 
available. 

We consider the calculation of the virtual photon spin asymmetry
in DIS of a charged leptonic probe from a hadronic target within
the model of Ref.\cite{Paris03}. The Hamiltonian in this model
is chosen as
\beq
\label{eqn:H}
H=\bfalp\cdot\bvec{p} + \frac{1+\beta}{2}\st r,
\eeq
where $\bfalp$ and $\beta$ are Dirac matrices in the standard
representation \cite{BjD}.  It describes 
a massless Dirac particle in a linear confining well.  The 
half-vector plus half-scalar structure of the confining potential 
is chosen for its spin symmetry \cite{Page:2000ij}
wherein spin-orbit doublets are degenerate. It is motivated by the 
relatively small spin-orbit splittings seen in meson spectra.  
Computations are
simple with this choice since the lower components of the wave 
function are not coupled by the potential.
The value of the string tension $\st$ is assumed to be 1 GeV/fm 
as indicated by the slopes of baryon Regge trajectories. 
In Ref.\cite{Paris03} all the eigenstates of this model were obtained 
exactly for excitation energies up to $\sim 12$ GeV. The ground 
state energy, $E_0$ for this string tension is 840 MeV.
The model may be viewed as a
heavy-light meson, such as $\bar{t}u$, in the limit 
that the antiquark mass goes to infinity.
However, it retains only the confining part of the $\bar{t}u$ 
interaction modeled by a flux tube. 

The model neglects gluon and sea-quark contributions to DIS
as well as the QCD evolution. However, the observed ratio of 
the $g_1(x)$ to $F_1(x)$, the unpolarized structure function
is relatively independent of $Q^2$ \cite{Filippone01}, and our 
objective is to calculate the $x$-dependence of this ratio for 
the contribution of valence quarks to DIS. We hope that the model 
is useful in this context. 

The virtual photon asymmetry is defined as \cite{Filippone01}
\beq
\label{eqn:A1}
A_1=\frac{\sigma_{\shalf}-\sigma_{\sthalf}}{\sigma_{\shalf}+\sigma_{\sthalf}}
\eeq
with $\sigma_{\shalf}$ and $\sigma_{\sthalf}$ the helicity cross sections
for the target angular momentum antiparallel and parallel to the photon
helicity, respectively. We may calculate the inclusive virtual 
photon helicity cross sections in the rest frame of the target as
\beqa
\label{eqn:sig1/2}
\sigma_{\shalf}(\qmag,\nu) &=& \sigma_M \sum_{I}
\left|\bra{I}\alpha_+ e^{i\qmag z}\ket{0,-{\shalf}}\right|^2
\nonumber \\
&\times&\delta(E_I-E_0-\nu) \\
\label{eqn:sig3/2}
\sigma_{\sthalf}(\qmag,\nu) &=& \sigma_M \sum_{I}
\left|\bra{I}\alpha_+ e^{i\qmag z}\ket{0,+{\shalf}}\right|^2
\nonumber \\
&\times&\delta(E_I-E_0-\nu)
\eeqa
where $\qmag$ and $\nu$ are the momentum and energy transferred to the 
target, $\sigma_M$ is the Mott cross section and we have assumed that the
virtual photon is in the $\hat{\bvec{z}}$ direction. The ground states 
$\ket{0,j_z={\scriptstyle\pm\shalf}}$ have the total angular momentum 
projection $j_z=\pm\half$. The operator $\alpha_+$ corresponds 
to a virtual photon with positive helicity, and $\ket{I}$ are 
eigenstates of the 
Hamiltonian $H$ [Eq.(\ref{eqn:H})] with energies $E_I$.  

\renewcommand{\bottomfraction}{0.95}
\begin{figure}[b]
\includegraphics[ width=230pt, keepaspectratio, angle=0, clip ]{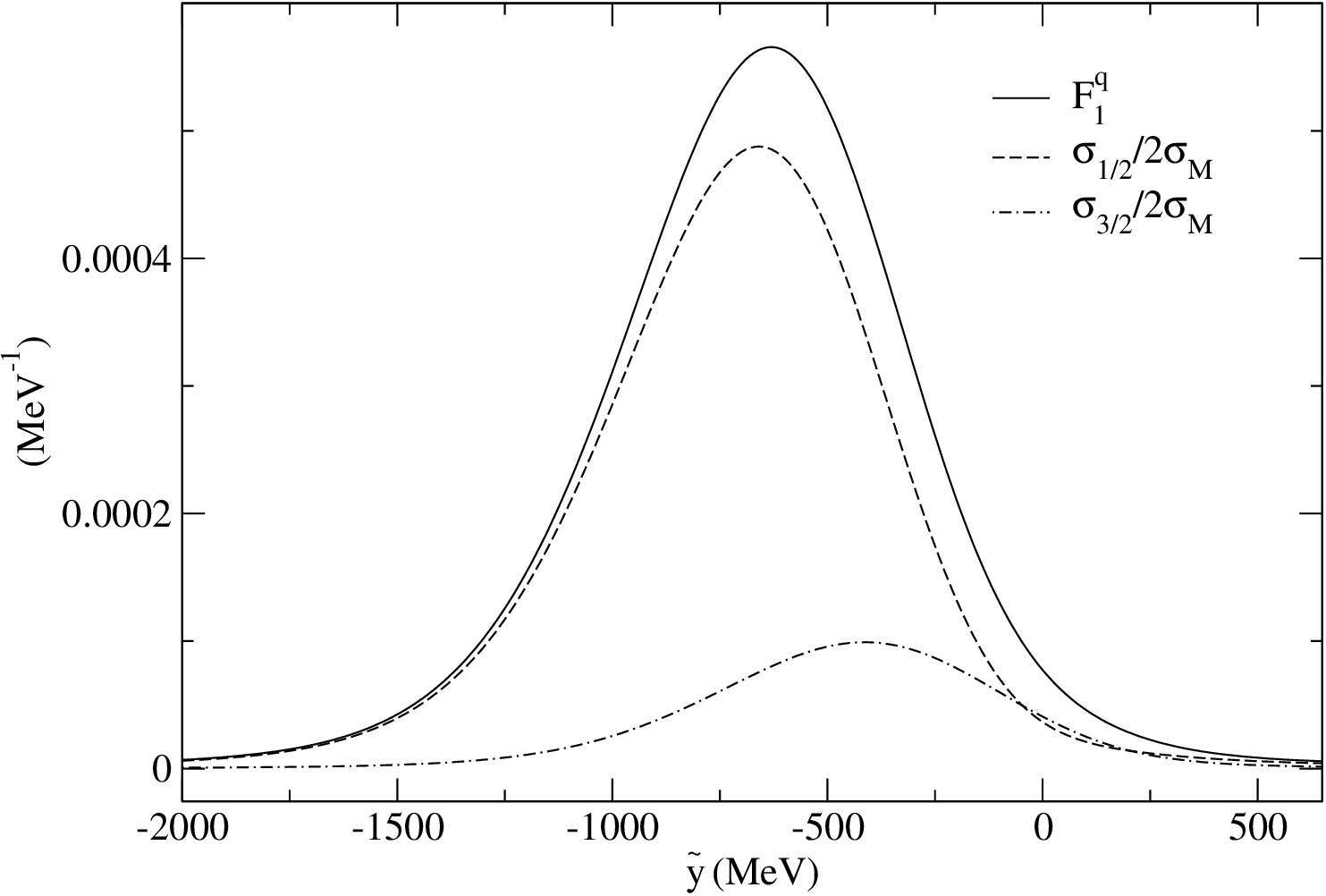}
\caption{\label{fig:hxs} Virtual photon helicity cross section of a confined 
massless quark, modulo
twice the Mott cross section, as a function of $\ysc$. 
The dashed $(\sigma_{\shalf})$ and dash-dotted $(\sigma_{\sthalf})$ curves sum to the
unpolarized structure function (solid curve).}
\end{figure}

The calculation of the virtual photon helicity cross sections
proceeds in this model, without approximations, exactly as the calculation 
of the unpolarized structure
functions described in Ref.\cite{Paris03}. When $\qmag$ is large 
the $\sigma/\sigma_M$ depend only on $\ysc=\qmag-\nu$.  Figure (\ref{fig:hxs})
shows the calculated $\sigma_{\shalf}/(2\sigma_M)$ 
and $\sigma_{\sthalf}/(2\sigma_M)$
plotted as a function of the scaling variable $\ysc$, 
and their sum
\beq
\label{eqn:F1q}
F^q_1(\ysc) = 
\frac{1}{2\sigma_M}\left(\sigma_{\shalf}+\sigma_{\sthalf}\right),
\eeq
the unpolarized structure function.
The conventionally used Bjorken and Nachtmann scaling variables 
are related to $\ysc$ by \cite{BPS00}: 
\beq
x (Q^2 \rightarrow \infty) = \xi = -\frac{\ysc}{M_T},
\label{eq:xyz}
\eeq
where $M_T$ is the target mass.
Thus small (large) negative $\ysc$ correspond to small (large) $x$. 
We note that the $\sigma_{\shalf}(\ysc)$ and $\sigma_{\sthalf}(\ysc)$ are not 
proportional, which implies that the $A_1^q$ of a confined relativistic quark 
has a large $\ysc$ or equivalently $x$ dependence. 

The ground state $|0,j_z\rangle$ of the confined quark is given by: 
\beqa
\label{fig:gswf}
\Psi_{0,j_z}(\rvec) &=& \left(
\begin{array}{c}
f_0(r)\ysa{\half}{j_z}{0}(\bvec{\hat{r}}) \\
ig_0(r)\ysa{\half}{j_z}{1}(\bvec{\hat{r}}) \end{array}\right),
\eeqa
where, $f_0(r)$ and $g_0(r)$ are the radial functions for the $s$--
and $p$--waves, respectively, and 
$\ysa{j}{j_z}{\ell}$ are the spin-angle functions obtained by 
coupling spin and orbital angular momentum to $j=\half$. 

\begin{figure}[b]
\includegraphics[ width=230pt, keepaspectratio, angle=0, clip ]{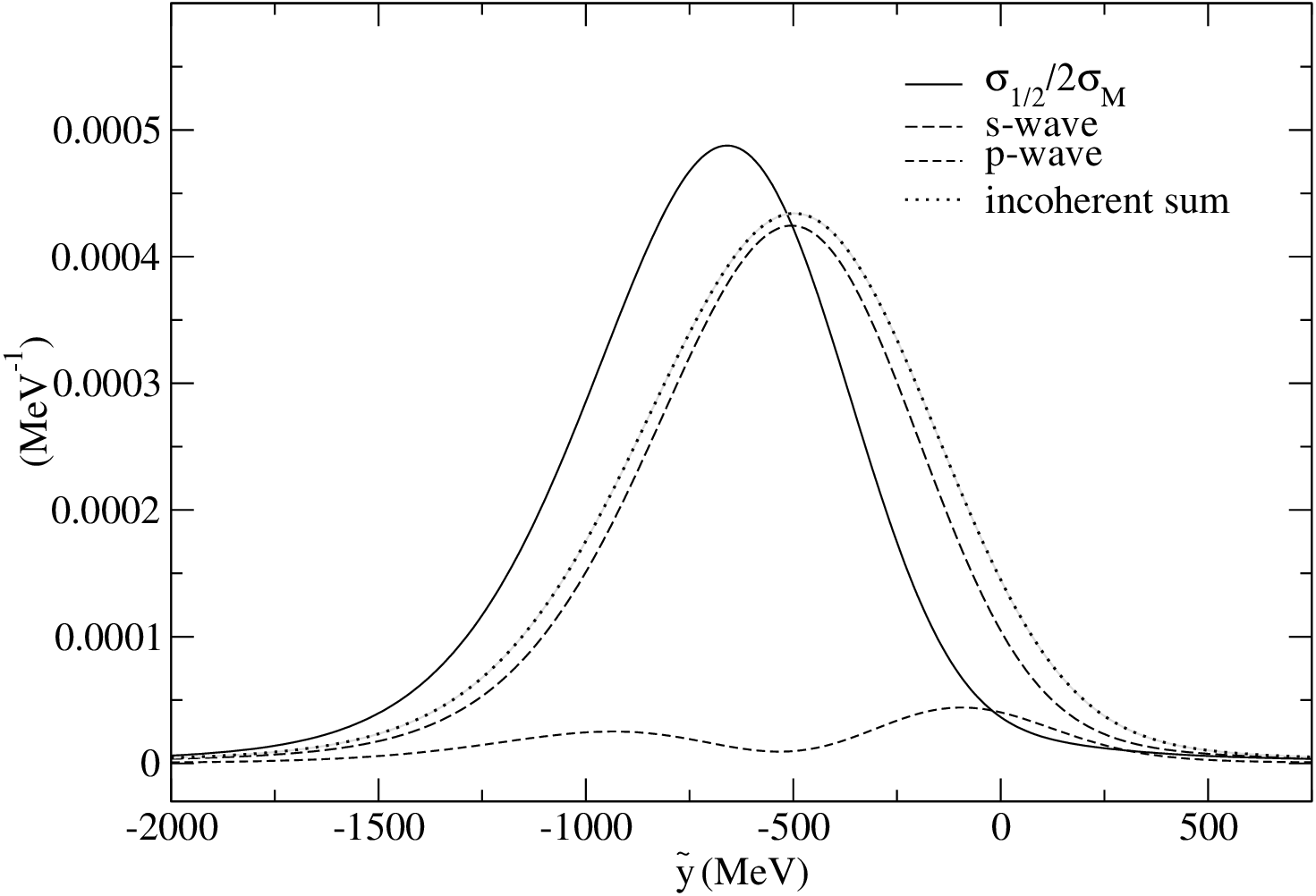}
\caption{\label{fig:intfr} Interference effects in $j_z=-\half$
($\sigma_\half$) structure function. The dashed lines give the contributions 
of the $s$-- and $p$--waves alone, the dotted line shows their incoherent sum 
and full line is the exact result. }
\end{figure}

The interference in the DIS between the $s$-- and $p$--waves 
contributes significantly to the $\ysc$ dependence of the 
$\sigma_{\shalf}$ helicity cross-section, $A_1^q$ and $F_1^q$. The effect
of interference is shown in Fig.(\ref{fig:intfr}) where we compare the 
polarized cross section $\sigma_\half$ including interference terms (solid
curve, labeled `full') with the polarized cross section neglecting 
interference terms (dotted curve). Also shown are the polarized cross 
sections obtained after keeping only the $s$-- or $p$--waves in the  
$j_z=-\half$ target. We note that the interference shifts 
$\sigma_\half$ to more 
negative $\ysc$ corresponding to larger values of $\xi$.  
Only the $p$--waves contribute to $\sigma_{\sthalf}$ shown 
in Fig.(\ref{fig:hxs}). 

The virtual photon asymmetry is given in terms of the spin-dependent
structure functions $g_1$ and $g_2$ \cite{Filippone01} by
\beq
\label{eqn:A1g1}
A_1=\frac{g_1 - \gamma^2 g_2}{F_1} \approx \frac{g_1}{F_1}
\eeq
where $\gamma^2=4M_T^2 x^2/Q^2$, in the scaling regime,
$Q^2\rightarrow\infty$.  As mentioned earlier, the observed 
$A_1$ of the proton, $A_1^p$ is largely independent of $Q^2$, 
and is used to extract values of $g_1^p/F_1^p$ \cite{Filippone01}. 

\begin{figure}
\includegraphics[ width=220pt, keepaspectratio, angle=0, clip ]{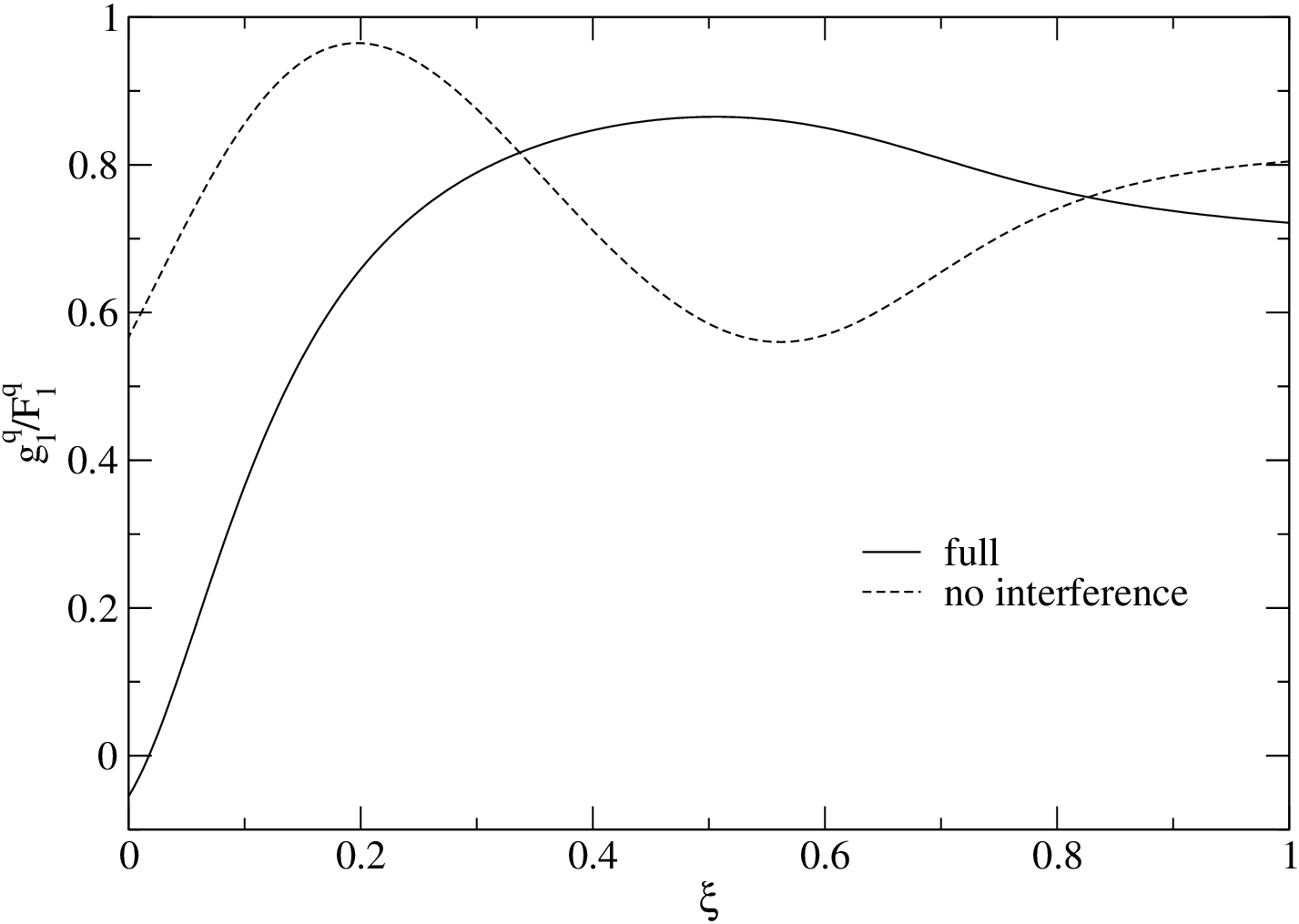}
\caption{\label{fig:g1of1} The $g^q_1/F^q_1$ for a single massless quark
confined by a flux-tube, as a function of the Nachtmann 
$\xi=(\qmag-\nu)/M_T$ with and without interference terms (see text).}
\end{figure}

Using the structure functions given in Fig.(\ref{fig:hxs})
we can easily calculate 
the virtual photon asymmetry $A^q_1$ or equivalently the ratio 
$g^q_1/F^q_1$ for a single confined quark, as a function of $\ysc$.  
In order to compare it with the data on protons we have to convert 
it to a function of $\xi$ by providing a mass scale $M_T$ (see 
Eq.(\ref{eq:xyz})).  Our model target has infinite mass associated with 
the center of the confining potential.  However, that mass is 
not relevant since only the confined quark contributes to DIS.  We 
use $M_T=2.5$ GeV $\sim 3E_0$, where $E_0$ is the energy of a single 
confined quark in the ground state.  With this choice the $F_1^q(\xi)$ 
becomes small at 
$\xi \sim 0.8$ as in the proton. The solid curve in Fig.(\ref{fig:g1of1}) 
shows  the $A_1^q(\xi)$ or equivalently $g_1^q(\xi)/F_1^q(\xi)$ of a 
confined quark. The calculated ratio goes to zero at small $\xi$, and 
this behavior is independent of the chosen value of $M_T$.  The dip 
at $\xi=0$ is due to the shift of $\sigma_\half$ to larger values of 
$\xi$, produced by the interference effect shown in Fig.(\ref{fig:intfr}). 
When the interference 
terms are omitted we obtain the dashed curve in Fig.(\ref{fig:g1of1}) 
which has $g_1^q/F_1^q \sim 0.6$ at $\xi=0$.  

Alternatively we could have chosen the string tension $\sqrt{\sigma}$ 
such that $3E_0=M_N$, the nucleon mass.  However, since $\sqrt{\sigma}$ 
provides the only mass scale in the Hamiltonian $H$ [Eq.(\ref{eqn:H})], 
this choice gives exactly the same $A_1^q(\xi)$ as the previous.  

In the remaining parts of this letter we attempt to estimate the $\xi$ 
dependence of $g_1^p/F_1^p$ ratio for the proton 
in the ``naive'' quark model in which the total PDF (parton distribution 
functions) are approximated by 
the sum of the three valence quark contributions.  The unpolarized PDF 
of valence quarks in $j_z=\pm \half$ states are denoted by 
$u(\pm {\shalf},\xi)$ and $d(\pm {\shalf},\xi)$.  We extract these from the 
MRST \cite{MRST02} PDF fitted to the experimental data, and use only 
the calculated asymmetry $A^q_1(\xi)$. 

In order to obtain the asymmetry of the proton we must account
for the flavor dependence of the PDF's, as extracted from experiment.
It is well known that the flavor dependence of the valence PDF's
can be understood in terms of the state of the residual system
after removal of the struck quark \cite{CT88,BPS00}. The residual
diquark in DIS from protons can be in a spin-0 or spin-1 state.
Removal of a $d$ quark results in a spin-1 diquark, while removal
of a $u$ quark results in a spin-0 diquark 3/4 of the time and
spin-1 for the remainder. 

Let $q_0(\xi)$ and $q_1(\xi)$ be the unpolarized PDF when the diquark is in 
spin-0 and 1 state respectively.  From the proton spin-flavor wave function 
we obtain
\beqa
\label{eqn:pdfs}
u(+{\shalf},\xi) &=& \frac{3}{2}q_0(\xi)+\frac{1}{6}q_1(\xi) \nonumber \\
u(-{\shalf},\xi) &=& d(+{\shalf},\xi) = \frac{1}{3}q_1(\xi) \nonumber \\
d(-{\shalf},\xi) &=& \frac{2}{3}q_1(\xi).
\eeqa
The empirically known, total unpolarized $u(\xi)$ and $d(\xi)$ are 
given by
\beqa
u(\xi) &=& u(+{\shalf},\xi)+u(-{\shalf},\xi)  \nonumber \\
&=&\frac{3}{2}q_0(\xi)+\half q_1(\xi) \nonumber \\ 
\label{eqn:qprot}
d(\xi) &=& d(+{\shalf},\xi)+d(-{\shalf},\xi) = q_1(\xi).
\eeqa
Solving the above equations for $q_0(\xi)$ and $q_1(\xi)$ using the MRST
\cite{MRST02} $u(\xi)$ and $d(\xi)$ 
we find that $q_0$ is shifted to larger values of $\xi$ with respect to 
$q_1$ as expected from the smaller mass of the spin-0 diquark. Eliminating 
$q_0$ and $q_1$ from Eqs.(\ref{eqn:pdfs}) and (\ref{eqn:qprot}) we obtain: 
\beqa
\label{eqn:pdfs2}
u(+{\shalf},\xi) &=& u(\xi)-\frac{1}{3}d(\xi) \nonumber \\
u(-{\shalf},\xi) &=& d(+{\shalf},\xi) = \frac{1}{3}d(\xi) \nonumber \\
d(-{\shalf},\xi) &=& \frac{2}{3}d(\xi).
\eeqa

Due to the presence of $p$--waves both the $j_z=\pm {\shalf}$ PDF's 
have spin $\uparrow$ and $\downarrow$ contributions.  For example the 
$u(j_z,\xi)$ are given by:
\beqa
\label{eqn:unpol}
u_\uparrow(\pm{\shalf},\xi)+u_\downarrow(\pm{\shalf},\xi)&=&u(\pm{\shalf},\xi) \\
\label{eqn:Aq1tm}
u_\uparrow(\pm{\shalf},\xi) - u_\downarrow(\pm{\shalf},\xi) &=& \pm A^q_1(\xi) u(\pm{\shalf},\xi),
\eeqa
where $A^q_1(\xi)$ is the spin asymmetry of a quark in the $j={\shalf}$ ground state. 
In the following we will use $A^q_1=g^q_1/F^q_1$ calculated above using linear 
confinement, and shown in Fig.(\ref{fig:g1of1}). 

The total spin $\uparrow,\downarrow$ PDF's, summed over $j_z$, are obtained 
from the above equations. They are given by:
\beqa
\label{eqn:ubpdfs}
u_{\uparrow,\downarrow}(\xi) &=& \frac{u(\xi)}{2}\left[1\pm A^q_1(\xi)\right] 
\mp \frac{d(\xi)}{3}A^q_1(\xi) \\
\label{eqn:dbpdfs}
d_{\uparrow,\downarrow}(\xi) &=& 
\frac{d(\xi)}{2}\left[1\mp \frac{1}{3}A^q_1(\xi)\right],
\eeqa

\begin{figure}[t]
\includegraphics[ width=220pt, keepaspectratio, angle=0, clip ]{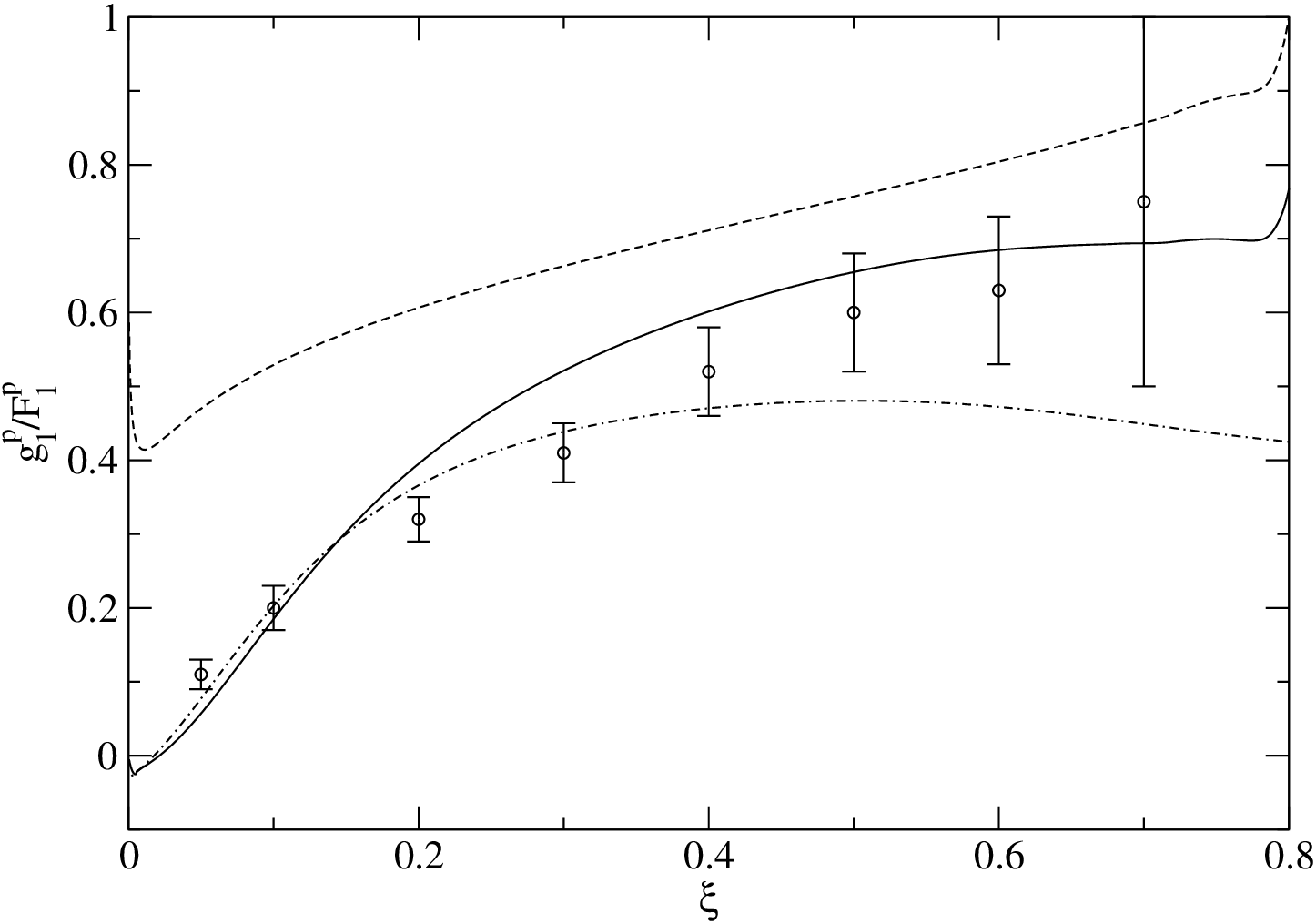}
\caption{\label{fig:g1of1N} The $g^p_1/F^p_1$ from MRST $u(\xi)$ and $d(\xi)$.
The full line shows results obtained with the calculated $A^q_1$ while the 
dashed line shows results assuming $A^q_1=1$. 
The dash-dot line is obtained using the approximation $u(\xi)=2d(\xi)$. 
The data is from Ref.\cite{Filippone01}.}
\end{figure}

We may now compute the spin asymmetry in the proton. Using
\beqa
\label{eqn:F1qpm}
F_1(\xi) &=& \half\left[
\frac{4}{9}u(\xi)
+\frac{1}{9}d(\xi) \right] \\
\label{eqn:g1qpm}
g_1(\xi) &=& \half\left[
\frac{4}{9}\left(u_\uparrow(\xi)-u_\downarrow(\xi)\right)
\right. \nonumber \\
&+&\left.
\frac{1}{9}\left(d_\uparrow(\xi)-d_\downarrow(\xi)\right)\right],
\eeqa
and obtain
\beq
\label{eqn:g1of1p}
\frac{g_1^p(\xi)}{F_1^p(\xi)}
=\left(\frac{4u(\xi)-3d(\xi)}{4u(\xi)+d(\xi)}\right)A_1^q(\xi).
\eeq
This ratio is plotted in Fig.(\ref{fig:g1of1N}) as the solid 
curve and is in fair agreement with the data from 
Ref.\cite{Filippone01} at all $\xi$.
The PDF's evolved to $Q^2=5$ GeV$^2$ at next-to leading order
are used to obtain the results 
shown in Fig.(\ref{fig:g1of1N}), but 
the ratio, $(4u(\xi)-3d(\xi))/(4u(\xi)-d(\xi))$ is fairly 
insensitive to $Q^2$ in the range $2 < Q^2 < 20$ GeV$^2$. 

It should be pointed out that at $\xi < 0.2$ the sea quark contributions 
to the $F_1^p(\xi)$ are large, particularly at large $Q^2$.  However,
they are neglected in the present calculation.

If we neglect $p$--waves in the valence quark orbitals, 
then $A^q_1(\xi)=1$.  The resulting $g^p_1/F^p_1$ is significantly above the 
experimental data as shown in Fig.(\ref{fig:g1of1N}). 
In the $SU(6)$ limit of the
naive quark model we have $u(\xi)=2d(\xi)$ and
Eq.(\ref{eqn:g1of1p}) reduces to 
\beq
\label{eqn:g1of1su6}
\frac{g_1^p(\xi)}{F_1^p(\xi)}
=\frac{5}{9}A_1^q(\xi).
\eeq
Results obtained with this approximation are shown by the dot-dashed curve 
in Fig.(\ref{fig:g1of1N}).  It lies below the data at large $\xi$.  
Approximating the $A^q_1$ by unity in the above equation gives the 
$\xi$ independent result, $g_1^p/F_1^p = 5/9$, of Kuti and 
Weisskopf \cite{Kuti71}. 

In conclusion the present work suggests that at least two different 
effects shape the $\xi$ dependence of $A_1^p(\xi)$.  The $p$--waves in 
bound quark wave functions interfere with the dominant $s$-waves to 
suppress $A_1^q(\xi)$ at small $\xi$; and the difference in the 
shapes of the $u(\xi)$ and $d(\xi)$ enhances the $A_1^p(\xi)$ at 
large $\xi$.  

Our model is certainly too simple; it approximates the 
problem of three interacting quarks by a relativistic one-quark 
problem.  Nevertheless $p$--waves occur very naturally in the 
wave functions of spin-half relativistic particles, and their effect 
will presumably exist in more refined treatments of spin asymmetries. 

\acknowledgments
This work has been supported by the US National Science Foundation 
via grant PHY-00-98353, and by the US Department of Energy 
under contract W-7405-ENG-36.
 
\bibliography{dissa}

\end{document}

%% file: defs.tex

\newcommand{\alps}{\ensuremath{\alpha_s}}
\newcommand{\qbar}{\bar{q}}
\newcommand{\beq}{\begin{equation}}
\newcommand{\eeq}{\end{equation}}
\newcommand{\beqa}{\begin{eqnarray}}
\newcommand{\eeqa}{\end{eqnarray}}
\newcommand{\gs}{g_{\pi NN}}
\newcommand{\gw}{f_\pi}
\newcommand{\mq}{m_Q}
\newcommand{\mn}{m_N}
\newcommand{\bb}{\langle}
\newcommand{\kb}{\rangle}
\newcommand{\st}{\ensuremath{\sqrt{\sigma}}}
\newcommand{\rvec}{\mathbf{r}}
\newcommand{\bvec}[1]{\ensuremath{\mathbf{#1}}}
\newcommand{\bra}[1]{\ensuremath{\bb#1|}}
\newcommand{\ket}[1]{\ensuremath{|#1\kb}}
\newcommand{\gft}{\ensuremath{\gamma_{FT}}}
\newcommand{\bfalp}{\mbox{\boldmath{$\alpha$}}}
\newcommand{\bfnab}{\mbox{\boldmath{$\nabla$}}}
\newcommand{\bfpi}{\mbox{\boldmath{$\pi$}}}
\newcommand{\bfsig}{\mbox{\boldmath{$\sigma$}}}
\newcommand{\bftau}{\mbox{\boldmath{$\tau$}}}
\newcommand{\spup}{\uparrow}
\newcommand{\spd}{\downarrow}
\newcommand{\hbarom}{\frac{\hbar^2}{m_Q}}
\newcommand{\half}{\frac{1}{2}}
\newcommand{\thalf}{\frac{3}{2}}
\newcommand{\shalf}{\scriptstyle\frac{1}{2}}
\newcommand{\sthalf}{\scriptstyle\frac{3}{2}}
\newcommand{\vnn}{\ensuremath{\hat{v}_{NN}}}
\newcommand{\argonne}{\ensuremath{v_{18}}}
\newcommand{\lqcd}{\ensuremath{\mathcal{L}_{QCD}}}
\newcommand{\lgf}{\ensuremath{\mathcal{L}_g}}
\newcommand{\lqm}{\ensuremath{\mathcal{L}_q}}
\newcommand{\lqg}{\ensuremath{\mathcal{L}_{qg}}}
\newcommand{\nn}{\ensuremath{NN}}
\newcommand{\hpnd}{\ensuremath{H_{\pi N\Delta}}}
\newcommand{\hpqq}{\ensuremath{H_{\pi qq}}}
\newcommand{\fpnn}{\ensuremath{f_{\pi NN}}}
\newcommand{\fpnd}{\ensuremath{f_{\pi N\Delta}}}
\newcommand{\fpqq}{\ensuremath{f_{\pi qq}}}
\newcommand{\ylm}{\ensuremath{Y_\ell^m}}
\newcommand{\ylmc}{\ensuremath{Y_\ell^{m*}}}
\newcommand{\qbh}{\hat{\bvec{q}}}
\newcommand{\xbh}{\hat{\bvec{X}}}
\newcommand{\dt}{\Delta\tau}
\newcommand{\qmag}{|\bvec{q}|}
\newcommand{\pmag}{|\bvec{p}|}
\newcommand{\oas}{\ensuremath{\mathcal{O}(\alpha_s)}}
\newcommand{\vtxb}{\ensuremath{\Lambda_\mu(p',p)}}
\newcommand{\vtxp}{\ensuremath{\Lambda^\mu(p',p)}}
\newcommand{\pwqp}{e^{i\bvec{q}\cdot\bvec{r}}}
\newcommand{\pwqm}{e^{-i\bvec{q}\cdot\bvec{r}}}
\newcommand{\gsa}[1]{\ensuremath{\bb#1\kb_0}}
\newcommand{\oer}[1]{\mathcal{O}\left(\frac{1}{\qmag^{#1}}\right)}
\newcommand{\nub}[1]{\overline{\nu^{#1}}}
\newcommand{\balph}{\mbox{\boldmath{$\alpha$}}}
\newcommand{\bgam}{\mbox{\boldmath{$\gamma$}}}
\newcommand{\epf}{E_\bvec{p}}
\newcommand{\epfp}{E_{\bvec{p}'}}
\newcommand{\eka}{E_{\alpha\kappa}}
\newcommand{\ekaq}{(E_{\alpha\kappa})^2}
\newcommand{\ekap}{E_{\alpha'\kappa}}
\newcommand{\ekpa}{E+{\alpha\kappa_+}}
\newcommand{\ekma}{E_{\alpha\kappa_-}}
\newcommand{\ekp}{E_{\kappa_+}}
\newcommand{\ekm}{E_{\kappa_-}}
\newcommand{\ekpap}{E_{\alpha'\kappa_+}}
\newcommand{\ekmap}{E_{\alpha'\kappa_-}}
\newcommand{\yjm}[1]{\mathcal{Y}_{jm}^{#1}}
\newcommand{\ysa}[3]{\mathcal{Y}_{#1,#2}^{#3}}
\newcommand{\ysc}{\tilde{y}}